# Dirac-graphene in slow-light

P.A. Golovinski[1,2], V.A. Astapenko[1], A.V. Yakovets[1]

[1] *Moscow Institute of Physics and Technology, Russia,*
[2] *Physics Research Laboratory, Voronezh Technical State University, Russia*
e-mail: golovinski@bk.ru

The unique properties of graphene are the subject of broad fundamental and applied research [1, 2]. Much of the interest in them is due to the fact that the quasi-particles in graphene have special properties, being massless fermions satisfying the two-dimensional Dirac equation [3, 4]. Since the Fermi velocity in graphene significantly less light speed in vacuum *c*, quantum electrodynamics' nonlinear processes will occur in graphene in fields of moderate intensity compared to the intensity of the superstrong relativistic fields that are required for the observation of nonlinear effects in vacuum [5]. This makes graphene a unique model system that allows for study the effects of the quantum electrodynamics of superstrong fields [6-8], as well as through the curved graphene the processes of general relativity, including the Hawking effect [9]. We obtained exact solutions for the wave function and the Green's function in the slow light pulse with the group velocity, consistent with the Fermi velocity in graphene.

The dynamics of graphene quasiparticles in an external plane wave field is described by the Dirac equation for massless fermions [4, 10]

$$i\frac{\partial \psi}{\partial \tau} = \boldsymbol{\sigma}(\mathbf{p} - e\mathbf{A}(x,y,t))\psi. \qquad (1)$$

Here, $\psi$ is two-component wave function

$$\psi = \begin{pmatrix} \psi_A \\ \psi_B \end{pmatrix}, \qquad (2)$$

$\mathbf{p} = \mathbf{e}_x \partial_x + \mathbf{e}_y \partial_y$, $\boldsymbol{\sigma} = \mathbf{e}_x \sigma_x + \mathbf{e}_y \sigma_y$, $\sigma_x, \sigma_y$ are Pauli matrixes. Variable $\tau$ in terms of time $t$ has the form $\tau = v_F t$, where $v_F$ is the Fermi velocity, and $\mathbf{q} = (x, y)$. We use the relativistic system of units, where $\hbar = c = 1$.

The field of a plane linearly polarized electromagnetic wave on the graphene surface is described by the equation

$$\mathbf{A}(\xi) = \mathbf{e}_x A(\xi), \xi = \alpha y - \tau, . \tag{3}$$

For a parameter value $\alpha = 1$ electromagnetic wave and graphene quasiparticles velocities coincide.

To solve the problem with the field introduce the notation

$$L^+ = h + i\partial_\tau, \; L^- = h - i\partial_\tau, \tag{4}$$

where

$$h = \begin{pmatrix} 0 & p_x - ip_y + a(y,\tau) \\ p_x + ip_y + a(y,\tau) & 0 \end{pmatrix}, a = -eA, \tag{5}$$

and Eq. (1) is simply $L^-\psi = 0$. It is clear, that is valid a quadrated equation

$$H\psi = 0 \tag{6}$$

with

$$H = L^+ L^- == \partial_\tau^2 - \partial_x^2 - \partial_y^2 - 2ia\partial_x + a^2 + a'\beta \tag{7}$$

and $\beta = \sigma_x + \alpha\sigma_z$.

Following the Volkov's method, we seek a solution of Eq. (6) in the form

$$\psi_\mathbf{k} = e^{i\chi} F(\xi), \; \chi = \mathbf{kq} - k\tau, \xi = \alpha y - \tau. \tag{8}$$

The final solution of the Eq. (1) can be written as

$$\psi_\mathbf{k} = \exp\left(i(\mathbf{kq} - k\tau) + \frac{i}{2}\int_0^{y-\tau} d\xi \frac{2ak + a^2 + a'\beta}{k - k_y}\right) u_\mathbf{k}. \tag{9}$$

The singularity in the Eq. (9) is similar to the singularity in the Volkov's solution for 3 + 1 space in the massless limit.

We obtain the Green's function $G(q,q_1)$ as a solution of equation

$$L^-G(q,q_1)=\delta(x-x_1)\delta(q-q_1), \qquad (10)$$

where $q=(y,\tau)$. The projection of the momentum $k_x$ is a conserved quantity and solution is found in the form

$$G(q,q_1)=\frac{1}{2\pi}\int_{-\infty}^{\infty}dk_x\, e^{ik_x(x-x_1)}g_{k_x}(y,t,y_1,t_1). \qquad (11)$$

The function $g_{k_x}$ obeys equation

$$L_x^-g_{k_x}=\delta(y-y_1)\delta(t-t_1) \qquad (12)$$

and

$$L_x^\pm=\begin{pmatrix}0 & k_x-ip_y+a(y,\tau)\\ k_x+ip_y+a(y,\tau) & 0\end{pmatrix}\pm i\partial_\tau. \qquad (13)$$

At this point, we use the proper time Fock-Schwinger method [11-13]. We define the propagator

$$g_{k_x}(y,t,y_1,t_1)=-iL_x^+\int_0^\infty ds\,\langle q|e^{-isH_x}|q_1\rangle, \qquad (14)$$

where

$$H_x=p_y^2-p_\tau^2+b \qquad (15)$$

and $b=(k_x+a)^2+a'\beta$.

Let further $p_1=-i\partial_y$, $p_2=-i\partial_\tau$, $q_1=y, q_2=\tau$, $p=(p_1,p_2)$, $q=(q_1,q_2)$. Differentiating the matrix element $\langle q|e^{-isH_x}|q_1\rangle$ with respect to variable $s$ one can write down

$$i\partial_s\langle q|e^{-isH_x}|q_1\rangle=\langle q|H_xe^{-isH_x}|q_1\rangle. \qquad (16)$$

By inserting the unit operator $I=e^{-isH_x}ye^{isH_x}$ in the right-hand side of the Eq. (16), we obtain

$$i\partial_s \langle q,s|q_1,0\rangle = \langle q,s|H_x(p(s),p(s))|q_1,0\rangle, \qquad (17)$$

where $|q,s\rangle$ $|q_1,0\rangle$ are the eigenvectors of operators $q_\mu(s)$ и $q_\mu(0)$ with eigenvalues $q$ and $q_1$: $q(s)|q,s\rangle = q|q,s\rangle$, $q(0)|q_1,0\rangle = q_1|q,0\rangle$, $p_\mu(s) = e^{isH_x} p_\mu e^{-isH_x}$, $q_\mu(s) = e^{isH_x} q_\mu e^{-isH_x}$. Variable $s$ is formally considered as a new "time". Operators $p(s) = (p_\xi(s), p_\eta(s))$, $q_\mu(s) = (\xi(s),\eta(s))$ in the Eq. (17) are taken in the Heisenberg representation, and satisfy the equations

$$\frac{dp_\mu(s)}{ds} = i[H_x, p_\mu(s)], \quad \frac{dq_\mu(s)}{ds} = i[H_x, q_\mu(s)]. \qquad (18)$$

With the help of the new variables $p_\eta = p_y - p_\tau, p_\xi = p_y + p_\tau$, $\xi = y - \tau, \eta = \tau + y$ the Hamiltonian $H_x$ can be written as

$$H_x = p_\eta p_\xi + b(y - \tau). \qquad (19)$$

The Heisenberg equations for the operators are of the form

$$\frac{dp_\xi(s)}{ds} = i[H_x, p_\xi(s)] = 0, \qquad (20)$$

$$\frac{dp_\eta(s)}{ds} = i[H_x, p_\eta(s)] = -2\frac{db}{d\xi},$$

$$\frac{d\xi(s)}{ds} = i[H_x, \xi(s)] = 2p_\xi(s),$$

$$\frac{d\eta(s)}{ds} = i[H_x, \eta(s)] = 2p_\eta(s).$$

After integrating the Heisenberg equations we have

$$\langle q,s|q_1,0\rangle = \frac{1}{s}\exp\left(\frac{i}{4s}\left((y-y_1)^2 - (\tau-\tau_1)^2\right) - \frac{is}{(y-y_1)-(\tau-\tau_1)}\int_{y_1-\tau_1}^{y-\tau} b\,d\xi\right). \qquad (21)$$

These solutions can be used as a basis when considering the various processes of interaction of ultrashort slow-light pulses with graphene. Such phenomena include, in particular, harmonic generation and impurity scattering.

The work was supported by the Ministry of Education of the Russian Federation GB № 2014/19.